\documentclass[sort&compress,final,3p,times,twocolumn,fixfloat]{elsarticle}

\usepackage{graphicx}% eps fig, package for more complicated commands
\usepackage{amssymb} % various useful mathematical symbols
\usepackage{amsmath} % extended theorem environments
\usepackage{comment}
\usepackage{lineno}
 \usepackage{ifpdf}
 \ifpdf
 \usepackage[pdftex]{hyperref}
 \else
 \usepackage[hypertex]{hyperref}
 \fi

 \hypersetup{
   pdftitle={},%
   pdfauthor={},%
   pdfsubject={},%
   pdfkeywords={},%
   pdfstartview={},%
   bookmarksopen=true, breaklinks=true, debug=true, %
   colorlinks=true, linkcolor=red, citecolor=blue, urlcolor=blue
 }

\usepackage{float} % improved interface for floating objects
\usepackage[caption=false]{subfig} % supersedes the subfigure package
\newcount\savefnused
\newcount\savefndone

\newcommand{\savefootnote}[2][\empty]% #1=number (optional), #2=text
{\ifx\empty#1\footnotemark\else\footnotemark[#1]\fi
 \global\advance\savefnused by 1
 \expandafter\xdef\csname savefnmark\the\savefnused\endcsname{\thefootnote}%
 \expandafter\xdef\csname savefntext\the\savefnused\endcsname{#2}%
}
\newcommand{\flushfootnote}{\loop\ifnum\savefndone<\savefnused
  \global\advance\savefndone by 1
  \footnotetext[\csname savefnmark\the\savefndone\endcsname]%
    {\csname savefntext\the\savefndone\endcsname}%
  \global\expandafter\let\csname savefnmark\the\savefndone\endcsname\relax
  \global\expandafter\let\csname savefntext\the\savefndone\endcsname\relax
\repeat}

\usepackage{multirow,bigdelim}
\usepackage{tabularx,ragged2e}%

\newcolumntype{Y}{>{\centering\arraybackslash}X}

\newcommand{\eqs}[1]{\begin{equation} \begin{split} #1\end{split} \end{equation} }

\newcommand{\nn}{\nonumber}

\def\ie{{\it i.e.}}
\def\eg{{\it e.g.}}

\def\GeV{{\rm GeV}}
\def\GeV2{{\rm GeV}^2}

\newcommand{\Q}{{\cal Q}}

\newcommand{\CS}{{\rm CS}}

\newcommand{\kTsqav}{\langle k_\sT^2 \rangle}

\newcommand{\mc}[1]{\mathcal{#1}}
\newcommand{\kT}{\bm{k}_{\sT}}
\newcommand{\koneT}{\bm{k}_{1\sT}}
\newcommand{\ktwoT}{\bm{k}_{2\sT}}
\newcommand{\qT}{\boldsymbol{P}_{\Q\Q\sT}}
\newcommand{\qTnorm}{{P_{\Q\Q}}_\sT}
\newcommand{\qTtwo}{{P^{2}_{\Q\Q}}_\sT}

\renewcommand{\d}{\mathrm{d}}
\newcommand{\sT}{{\scriptscriptstyle T}}

\newcommand{\ce}[1]{Eq.~(\ref{#1})}
\newcommand{\cf}[1]{{Fig.~\ref{#1}}}

\newcommand{\bm}[1]{\mbox{\boldmath $#1$}}

\begin{document} 
\begin{frontmatter}

\title{Pinning down the linearly-polarised gluons inside unpolarised protons using quarkonium-pair production at the LHC}

\author[a]{Jean-Philippe~Lansberg}
\address[a]{IPNO, CNRS-IN2P3, Univ. Paris-Sud, Universit\'e Paris-Saclay, 91406 Orsay Cedex, France}
\author[b,c]{Cristian Pisano}
\address[b]{Dipartimento di Fisica, Universit\`a di Pavia, and INFN, Sezione di Pavia
    Via Bassi 6, I-27100 Pavia, Italy}
\address[c]{Dipartimento di Fisica, Universit\`a di Cagliari, and INFN, Sezione di Cagliari
    Cittadella Universitaria, I-09042 Monserrato (CA), Italy}
\author[a,d]{Florent Scarpa}
\address[d]{Van Swinderen Institute for Particle Physics and Gravity,
University of Groningen, Nijenborgh 4, 9747 AG Groningen, The Netherlands}
\author[e,f]{Marc Schlegel}
\address[e]{Institute for Theoretical Physics, Universit\"{a}t T\"{u}bingen,  Auf der Morgenstelle 14, D-72076 T\"{u}bingen, Germany}
\address[f]{Department of Physics, New Mexico State University, Las Cruces, NM 88003, USA}
%\begin{linenumbers}

\begin{abstract}\small
{We show that the production of $J/\psi$ or $\Upsilon$ pairs in unpolarised $pp$ collisions is currently the best process to measure the momentum distribution of linearly-polarised gluons inside unpolarised
protons through the study of azimuthal asymmetries. Not only the short-distance coefficients for such reactions induce the largest possible $\cos 4\phi$ modulations, but analysed data are already available. Among  the various final states previously studied in unpolarised $pp$ collisions within the TMD approach, di-$J/\psi$ production exhibits by far the largest asymmetries, up to 50\% in the region studied by the ATLAS and CMS experiments. In addition, we use the very recent LHCb data at 13 TeV to perform the first fit of the unpolarised transverse-momentum-dependent gluon distribution.
}
\end{abstract}
\end{frontmatter}

%%%%%%%%%%%%%%%%%%%%%%%%%%%%%%%%%%%%%%%%%%%%%%%%%%%%%%%%%%%%%%%%%%%%%%%%%%%%
\section{Introduction}\vspace*{-0.2cm}
Probably one of the most striking phenomena arising from the extension of the collinear
factorisation --inspired from Feynman's and Bjorken's parton model-- to Transverse Momentum Dependent
(TMD) factorisation~\cite{Collins:2011zzd,Aybat:2011zv,GarciaEchevarria:2011rb,Angeles-Martinez:2015sea} is the appearance of azimuthal modulations induced by the polarisation of partons with nonzero
transverse momentum --even inside unpolarised hadrons. In the case of gluons in a proton, which trigger most 
of the scatterings at high energies, this new dynamics is encoded in the distribution $h_1^{\perp\,g}(x,\kT^2,\mu)$ 
of linearly-polarised gluons~\cite{Mulders:2000sh}. In practice, they generate $\cos 2\phi$ ($\cos 4\phi$) modulations in 
gluon-fusion scatterings where single (double) gluon-helicity flips occur. They can 
also alter transverse-momentum spectra, such as that of a 
$H^0$ boson~\cite{Boer:2011kf,Boer:2013fca}, via double gluon-helicity flips. 

In this Letter, we show that di-$J/\psi$ production, which among the 
quarkonium-associated-production processes has been the object of the 
largest number of experimental studies at the LHC and the Tevatron~\cite{Aaij:2011yc,Abazov:2014qba, Khachatryan:2014iia,Aaboud:2016fzt,Aaij:2016bqq}, is in fact the ideal process to perform the first measurement 
of $h_1^{\perp\,g}(x,\kT^2,\mu)$. It indeed exhibits the largest possible azimuthal asymmetries 
in regions already accessed by the ATLAS and CMS experiments 
where such modulations can be measured. Along the way of our study, we
perform the first extraction of $f_1^{g}(x,\kT^2,\mu)$ --its unpolarised
counterpart-- using recent LHCb data.

\section{TMD factorisation for gluon-induced scatterings}\vspace*{-0.2cm}
TMD factorisation extends collinear factorisation by accounting
for the parton transverse momentum, generally denoted by $\kT$.  {It applies to processes in which a momentum transfer is much larger than  $\vert \kT \vert $,  for instance at the LHC when a pair of particles ({\it e.g.}\  two quarkonium states $\Q$) is produced with a large invariant mass ($M_{\Q\Q}$)  as compared to its transverse momentum ($\qTnorm$).}
 
%%%%%%%%%%%%%%%%%%%%%%%%%%%%%%%%%%%%%%%%%%%%%%%%%%%%%%%%%%%%%%%%%%%%%%%%%%%%
\begin{figure}[hbt!]
\centering
\includegraphics[width=\columnwidth]{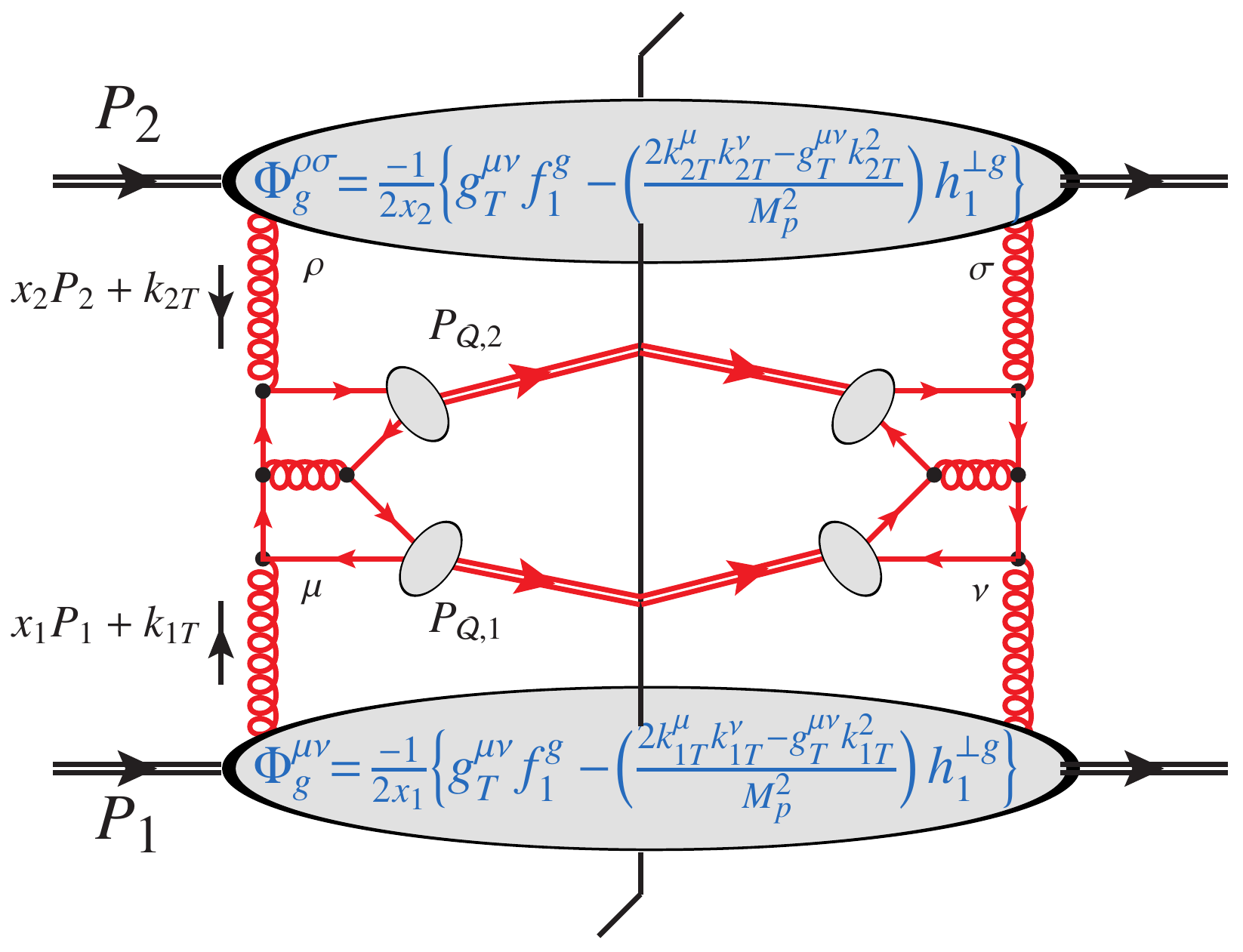}
\vspace*{-0.5cm}
\caption{Representative Feynman diagram for $p(P_1)\, {+}\, p(P_2) \to {\Q} (P_{\Q,1}) \,{+}  {\Q} (P_{\Q,2}) \,{+}X$ via 
gluon fusion at LO in the TMD framework.
}
\label{fig:Feynman-graph}
\end{figure}
%%%%%%%%%%%%%%%%%%%%%%%%%%%%%%%%%%%%%%%%%%%%%%%%%%%%%%%%%%%%%%%%%%%%%%%%%%%

{In practice, the gluon TMDs in an unpolarised  proton with a momentum $P$ and mass $M_p$ are defined through the hadron correlator~$\Phi_g^{\mu\nu}(x,\kT,\mu)$~\cite{Mulders:2000sh,Meissner:2007rx,Boer:2016xqr}, 
parametrised in terms of two independent TMDs,	the unpolarised distribution $f_1^g(x,\kT^2,\mu)$ and the distribution of linearly-polarised gluons $h_1^{\perp\,g}(x,\kT^2,\mu)$ (see~\cf{fig:Feynman-graph}),
where the gluon four-momentum $k$ is decomposed as $k = xP + k_{\sT} + k^-n$ [$n$ is any light-like vector ($n^2=0$) such that $n\cdot P \neq 0$], $\bm k_\sT^2 = -k_\sT^2$ and  $g^{\mu\nu}_{\sT} = g^{\mu\nu} - (P^{\mu} n^{\nu}+ P^\nu n^\mu)/P{\cdot} n$ and $\mu$ is the factorisation scale.

In the TMD approach and up to corrections suppressed by powers of the  observed system transverse momentum
over its invariant mass, the cross section for any gluon-fusion process (here $g(k_1) \,{+} \,g(k_2) \to  \Q(P_{\Q,1})\, {+} \, \Q(P_{\Q,2})\,$)
can be expressed as a contraction and a convolution
of a partonic short-distance contribution, $\cal M^{\mu\rho}$, with two gluon TMD correlators
evaluated at $(x_1,\koneT,\mu)$ and $(x_2,\ktwoT,\mu)$.
$\cal M^{\mu\rho}$ is simply calculated in perturbative QCD through a series expansion 
in $\alpha_s$~\cite{Ma:2012hh} using Feynman graphs (see~\cf{fig:Feynman-graph}).

Owing to process-dependent Wilson lines in the definition of the correlators
 which they parametrise, the TMDs are in general not universal. Physics wise, 
these Wilson lines describe the non-perturbative interactions of the active parton --the gluon in our case-- 
with soft spectator quarks and gluons in the nucleon  before or after 
the hard scattering. For the production of di-leptons, $\gamma\gamma$, di-$\Q$ or boson-$\Q$ pairs via a Color-Singlet (CS) transitions~\cite{Chang:1979nn,Baier:1981uk,Baier:1983va} -- \ie\ for purely colorless final states-- in $pp$ collisions,
only initial-state interactions (ISI) between the active gluons and the spectators can occur. 
Mathematically, these ISI can be encapsulated~\cite{Collins:2002kn} in TMDs  with 
past-pointing Wilson lines --the exchange can only occur before the hard scattering.
Such gluon TMDs correspond to the Weizs\"acker-Williams 
distributions relevant for the low-$x$ region \cite{Dumitru:2015gaa,Dominguez:2011wm}.
 
Besides, in lepton-induced production of colourful final states, like heavy-quark pair, dijet or $J/\psi$ (via Colour Octet (CO) transitions or states) production \cite{Boer:2010zf,Boer:2016fqd,Rajesh:2018qks}, to be studied at a future Electron-Ion Collider (EIC) \cite{Accardi:2012qut}, only final-state interactions (FSI) take place. 
Yet, since $f_1^g$ and $h_1^{\perp\,g}$ are time-reversal symmetric ($T$-even)\footnote{unlike other TMDs~\cite{Boer:1997nt,Boer:2003cm} such as the gluon distribution in a transversally polarised proton, also called the Sivers function~\cite{Sivers:1989cc}.}, TMD factorisation tells us 
that one in fact probes the same distributions
in both the production of {\it colourless} systems in hadroproduction with ISI and of {\it colourful} systems in leptoproduction with FSI.
In particular, one expects (see ~\cite{Boer:2016bfj} for further dicussions) that, 
\eqs{
f_1^{g\,[\gamma^\star p\to Q \bar Q X]}(x,\kT^2,\mu)&=f_1^{g\,[pp\to \Q \Q X]}(x,\kT^2,\mu),\\
h_1^{\perp,g\,[\gamma^\star p\to Q \bar Q X]}(x,\kT^2,\mu)&=h_1^{\perp,g\,[pp\to \Q \Q X]}(x,\kT^2,\mu).
}
In practice, this means that one should measure these processes at similar scales, $\mu$. The virtuality of the off-shell photon, $Q$, should be comparable to the invariant mass of the quarkonium pair, $M_{\Q\Q}$. If it is not the case, the extracted functions should be evolved to a common scale before comparing them.

Extracting these functions in different reactions is essential to test this 
universality property of the TMDs --akin to the well-known sign change of the quark Sivers effect~\cite{Brodsky:2002rv,Collins:2002kn}--, in order to validate TMD factorisation.

\section{Di-$\Q$ production \& TMD factorisation}\vspace*{-0.2cm}
For TMD factorisation to apply, di-$\Q$ production should at least satisfy both following conditions. 
First, it should result from a Single-Parton Scattering (SPS). Second, FSI should be negligible, which
is satisfied when quarkonia are produced via CS transitions~\cite{Ma:2012hh}. For completeness, we note that a formal proof of factorisation for such processes is still lacking. We also note that, in some recent works~\cite{Boer:2014lka,Mukherjee:2016cjw, DAlesio:2017rzj}, TMD factorisation has been assumed in the description of processes in which both ISI and FSI are present. In that regard, as we discuss below, the processes which we consider here are safer. 

The contributions of Double-parton-scatterings (DPSs) leading to di-$J/\psi$ 
is below 10\% for $\Delta y \sim 0$ in the CMS and ATLAS samples~\cite{Lansberg:2014swa,Aaboud:2016fzt}, that is away 
from the threshold with a $P_{\Q T}$ cut. In such a case, DPSs only become significant 
at large $\Delta y$.
In the LHCb acceptance, they cannot be neglected but can be
subtracted~\cite{Aaij:2016bqq} assuming the $J/\psi$ from DPSs to be uncorrelated; 
this is the standard procedure at LHC energies~\cite{Akesson:1986iv,Alitti:1991rd,Abe:1993rv,Abe:1997xk,Abazov:2009gc,Aad:2013bjm,Chatrchyan:2013xxa}.

 The CS dominance to the SPS yield is expected since each CO transition goes along with a relative suppression on the order
of $v^4$~\cite{Bodwin:1994jh,Cho:1995ce,Cho:1995vh} (see~\cite{Andronic:2015wma,Brambilla:2010cs,Lansberg:2006dh} for reviews) --$v$ being the heavy-quark velocity in the $\Q$ rest frame. For di-$J/\psi$ production
with $v^2_c\simeq 0.25$, the CO/CS yield ratio, scaling as $v_c^8$, is expected to be below the per-cent level since both the CO and the CS yields
appear at same order in $\alpha_s$, \ie\ $\alpha_s^4$. This has been corroborated by explicit computations~\cite{Ko:2010xy,Li:2013csa,Lansberg:2014swa} with corrections
from the CO states below the per-cent level in the region relevant for our study. Only in regions where DPSs 
are anyhow dominant (large $\Delta y$)~\cite{Lansberg:2014swa,He:2015qya,Baranov:2015cle} 
such CO contributions might become non-negligible because of specific kinematical enhancements~\cite{Lansberg:2014swa} which are however
irrelevant where we propose to measure di-$J/\psi$ production as a TMD probe.
We further note that the di-$J/\psi$ CS yield has been studied up to next-to-leading (NLO) 
accuracy in $\alpha_s$~\cite{Lansberg:2013qka,Sun:2014gca,Likhoded:2016zmk} in collinear factorisation.
The feed down from excited states is also not problematic for TMD factorisation to apply: $J/\psi+\chi_c$ production is suppressed~\cite{Lansberg:2014swa}
and $J/\psi+\psi'$ can be treated exactly like $J/\psi+J/\psi$.
For di-$\Upsilon$, the CS yield should be even more dominant and the DPS/SPS ratio should be small.   

Following~\cite{Lansberg:2017tlc}, the structure of the TMD cross section 
for $\Q\Q$ production reads
\begin{align}\label{eq:crosssection} 
&\frac{\d\sigma}{\d M_{\Q\Q} \d Y_{\Q\Q} \d^2 \qT \d \Omega} 
  =\frac{\sqrt{Q^2 - 4 M_\mc{Q}^2}}{(2\pi)^2 8 s\, Q^2}\,\Bigg\{
  F_1\, \mc{C} \Big[f_1^gf_1^g\Big] \nn \\& +
  F_2\, \mc{C} \Big[w_2h_1^{\perp g}h_1^{\perp g}\Big]+ 
\cos2\phi_{\CS} \Bigg(F_3 \mc{C} \Big[w_3 f_1^g h_1^{\perp g}\Big] \nn \\ & + F'_3 \mc{C} \Big[w'_3 h_1^{\perp g} f_1^g \Big]\Bigg)  + \cos 4\phi_{\CS}F_4 \mc{C}\! \left[w_4 h_1^{\perp g}h_1^{\perp g}\right]\!\Bigg \}\,,
\end{align}
where $\d\Omega=\d\!\cos\theta_{\CS}\d\phi_{\CS}$, $\{\theta_{\CS},\phi_{\CS}\}$ are the Collins-Soper (CS) angles~\cite{Collins:1977iv} and
$Y_{\Q\Q}$  is the pair rapidity -- $\qT$ and $Y_{\Q\Q}$ are defined in the hadron c.m.s. In the {\CS} frame, the $\Q$ direction is along $\vec e=(\sin\theta_{\CS}\cos \phi_{\CS},
\sin\theta_{\CS}\sin \phi_{\CS}, \cos\theta_{\CS})$. The overall factor is 
specific to the mass of the final-state particles and the analysed differential cross sections, and
the hard factors $F_i$ depend neither on $Y_{\Q\Q}$ nor on $\qT$. In addition, let us note
that --away from threshold-- $\cos\theta_{\CS} \sim 0$ corresponds to $\Delta y \sim 0$ in the hadron c.m.s., that is our preferred
region to avoid DPS contributions.
The TMD convolutions in~\ce{eq:crosssection} are defined as
\begin{multline}\label{eq:Cwfg}
\mathcal{C}[w\, f\, g] \equiv \int\!\! \d^{2}\koneT\!\! \int\!\! \d^{2}\ktwoT\,
  \delta^{2}(\koneT+\ktwoT-\qT)\\
\times  w(\koneT,\ktwoT)\, f(x_1,\koneT^{2},\mu)\, g(x_2,\ktwoT^{2},\mu) \, ,
\end{multline}
where $w(\koneT,\ktwoT)$ are generic transverse weights and $x_{1,2} =  \exp[\pm Y_{\Q\Q}]\, M_{\Q\Q}/\sqrt{s}$, with $s = (P_1 + P_2)^2$. 
The weights in \ce{eq:crosssection} are identical
for all the gluon-induced processes and can be found in~\cite{Lansberg:2017tlc}.

\section{The short-distance coefficients $F_i$}\vspace*{-0.2cm}
The factors $F_i$ are calculable process by process and we refer to~\cite{Lansberg:2017tlc} 
for details on how to obtain them from the helicity amplitudes. As such, they can be derived from the uncontracted 
amplitude given in~\cite{Qiao:2009kg}. For any process, $F^{(')}_{2,3,4} \leq F_1$. 
For $\Q\Q$ production,  they read 
\begin{align}%\label{eq:F1}
F_{1} &=  \frac{{\cal N}}{{\cal D} M_{\Q}^{2}} \sum^6_{i=0} f_{1,n}\ (\cos\theta_{\CS})^{2n}, \nn
\end{align}
\begin{align}
F_{2}  & =  \frac{2^4  3  M_{\Q}^{2}{\cal N}}{{\cal D} M_{\Q\Q}^4} \sum^4_{n=0} f_{2,n}\ (\cos\theta_{\CS})^{2n}, \nn\\
F'_{3}  &=F_{3}  =  \frac{- 2^3 (1-\alpha^{2}) {\cal N}}{{\cal D} M_{\Q\Q}^2} \sum^5_{n=0} f_{3,n}\ (\cos\theta_{\CS})^{2n},\nn\\
F_{4}&= \frac{(1-\alpha^{2})^2 {\cal N} }{{\cal D} M_{\Q}^{2}} \sum^6_{n=0} f_{4,n}\ (\cos\theta_{\CS})^{2n},
\end{align}
with $\alpha=2M_{\Q}/M_{\Q\Q}$, ${\cal N}={2^{11}3^{-4}\pi^{2}\alpha_{s}^{4}|R_\Q(0)|^{4}}$, ${\cal D}=M_{\Q\Q}^{4}\left(1-(1-\alpha^{2})\, c_{\theta}^{2}\right)^{4}$ and where $R_\Q(0)$ is the $\Q$ radial wave function at the origin. Note that the expressions are symmetric about $\theta_{\CS}=\pi/2$ since the process is forward-backward symmetric. The coefficient $f_{i,n}$ which are simple polynomials in $\alpha$ are given in the \ref{sec:appendix}. Like in collinear factorisation,  the  Born-order cross section scales as $\alpha_{s}^{4}$.

Both large and small $\Q\Q$ mass, $M_{\Q\Q}$,  limits are very interesting.  
Indeed, when $M_{\Q\Q}$ becomes much larger than the quarkonium mass, $M_{\Q}$, one finds that, for $\cos\theta_{\CS} \to 0$,    
\eqs{\label{eq:Fi_large_M}
&F_4 \to  F_1 \to  \frac{
%2^8 
256 {\cal N}}{M_{\Q\Q}^{4} M_{\Q}^{2}},}
\eqs{
F_{2} \!\to\! \frac{
%-3^4 \times 2^7 
81 M_{\Q}^{4}\cos\theta_{\CS}^{2}}{2 M_{\Q\Q}^4} \times F_1,}
\eqs{
&F_3 \!\to\!  \frac{- 
%3 \times 2^{11}
24 M_{\Q}^{2}\cos\theta_{\CS}^{2}}{M_{\Q\Q}^2} \times F_1.}

One first observes that $F_4 \to F_1$, for $\cos\theta_{\CS} \to 0$ 
away from the threshold --where the CMS and ATLAS data lie. 
This is the most important result of this study and is, to the best of our knowledge,
a unique feature of di-$J/\psi$ and di-$\Upsilon$ production. From this, it readily follows
that, for a given magnitude of $h_1^{\perp g}$, these processes will exhibit
the largest possible $\cos 4 \phi_{\CS}$ modulation, thus the highest possible sensitivity on
$h_1^{\perp g}$.

One also observes that $F_2$ ($F_3$) 
scales like $M^{-4}_{\Q\Q}$ ($M^{-2}_{\Q\Q}$) relative to $F_1$ and $F_4$. In other words,
the modification of the $\qTnorm$ dependence due to the linearly-polarised gluons encoded in $F_2$
vanishes at large invariant masses. In fact, it is also small at threshold,  $M_{\Q\Q} \to 2 M_{\Q}$, where one gets: 
\eqs{\label{eq:Fi_threshold} F_1 \to \frac{787{\cal N}}{16 M_{\Q}^{6}}, \quad F_2 \to \frac{3F_1}{787} , \quad  F_{3,4} \to 0.} 
$F_2$ can thus be neglected for all purposes in what follows.

Going back to the case where $M_{\Q\Q}^2 \gg 4 M^2_{\Q}$, the mass scaling in \ce{eq:Fi_large_M} also indicates that the $\cos 4 \phi_{\CS}$ modulation (double helicity flip) quickly takes over
the $\cos 2 \phi_{\CS}$ one (single helicity flip) and the $\cos\theta_{\CS}$ dependence 
indicates that $F_{2,3}$ are suppressed near $\Delta y\sim 0$.

As such, and thanks to the collected di-$J/\psi$ data, 
we conclude that this process is indeed the ideal one to extract
the linearly-polarised gluon distributions. The previously studied 
 $\gamma\gamma$~\cite{Qiu:2011ai}, $H^0$+jet~\cite{Boer:2014lka}, $\Q+\gamma$~\cite{Dunnen:2014eta},
$\Q+ \gamma^\star$ or
$\Q+Z$~\cite{Lansberg:2017tlc} processes show significantly smaller values of $F_4/F_1$, 
thus a strongly reduced sensitivity on  $h_1^{\perp g}$.

Knowing the $F_i$ and an observed differential yield, one can thus extract 
the various TMD convolutions of \ce{eq:Cwfg} from their azimuthal (in)dependent parts.
When the cross section is integrated over $\phi_{\CS}$,
the contribution from $F_{3,4}$ drops out from \ce{eq:crosssection} and only depends
on $\mc{C} \Big[f_1^gf_1^g\Big]$ and $\mc{C} \Big[w_2h_1^{\perp g}h_1^{\perp g}\Big]$.
To go further, we define   
$\cos n\phi_{\CS}$ [for $n=2,4$] weighted differential cross sections
normalised to the azimuthally independent term as: 
\eqs{\!\!\!\!\langle  \cos n\phi_{\CS} \rangle =
\frac{\int \!\!d\phi_{\CS} \cos n\phi_{\CS}\,  \frac{d\sigma}{d M_{\Q\Q} d Y_{\Q\Q} d^2 \qT d \Omega}}
{\!\!\int \!\!d\phi_{\CS} \frac{d\sigma}{d M_{\Q\Q} d Y_{\Q\Q} d^2\qT d \Omega}}.}
It is understood that $\langle \cos n\phi_{\CS} \rangle$ computed in a range of
$M_{\Q\Q}$, $Y_{\Q\Q}$, $\qTnorm$ or $\cos \theta_{\CS}$ is the ratio of corresponding integrals.
Using \ce{eq:crosssection},
one gets in a single phase-space point: 
\begin{align}
2\langle\cos 2\phi_{\CS}\rangle & = \frac{F_3 \mc{C} \Big[w_3 f_1^g h_1^{\perp g}\Big] + F'_3 \mc{C} \Big[w'_3 h_1^{\perp g} f_1^g \Big]}{F_1\, \mc{C} \Big[f_1^gf_1^g\Big]+F_2\, \mc{C} \Big[w_2h_1^{\perp g}h_1^{\perp g}\Big]}, \nn\\
2\langle\cos 4\phi_{\CS}\rangle & =  \frac{F_4 \mc{C}\! \left[w_4 h_1^{\perp g}h_1^{\perp g}\right]}{F_1\, \mc{C} \Big[f_1^gf_1^g\Big]+F_2\, \mc{C} \Big[w_2h_1^{\perp g}h_1^{\perp g}\Big]}.
\end{align}

\section{The transverse-momentum spectrum}\vspace*{-0.2cm}
Before discussing the expected size of the azimuthal asymmetries, 
let us have a closer look at the transverse-momentum dependence
of~\ce{eq:crosssection}, entirely encoded in $\mathcal{C}[w\, f\, g]$, which are process-independent, unlike the $F_i$. Since the gluon TMDs are still unknown,  we need to resort to models.

Following \cite{Schweitzer:2010tt}, one can assume a simple Gaussian dependence 
on $\kT^2$ for $f_1^g$, namely
\eqs{
f_1^g(x,\bm k_\sT^2,\mu) = \frac{g(x,\mu)}{\pi \langle  k_\sT^2 \rangle}\,
\exp\Big(-\frac{\bm k_\sT^2}{\langle  k_\sT^2 \rangle}\Big),\label{eq:Gaussf1}
}
where $g(x)$ is the collinear gluon PDF and $\langle  k_\sT^2 \rangle$ implicitly depends on the scale $\mu$.

Since $F_2$ is always small compared to $F_1$, the $\qTnorm$ spectrum in practice follows from the TMD 
convolution $\mathcal{C}[f_1 f_1]$ which only depends on $\langle  k_\sT^2 \rangle$. Conversely, 
one can thus fit $\langle  k_\sT^2 \rangle$ from the $\qTnorm$ spectrum recently measured 
by the LHCb Collaboration at 13~TeV~\cite{Aaij:2016bqq}  (see \cf{fig:f1_fit})
 from which we have the subtracted the DPS contributions evaluated by LHCb.
Such DPSs  are indeed expected to yield a different $\langle  \qTtwo \rangle$ since they
result from the convolution of two independent $2\to2$ scatterings.

\begin{figure}[hbt!]
\centering
{\includegraphics[height=\columnwidth,angle=-90]{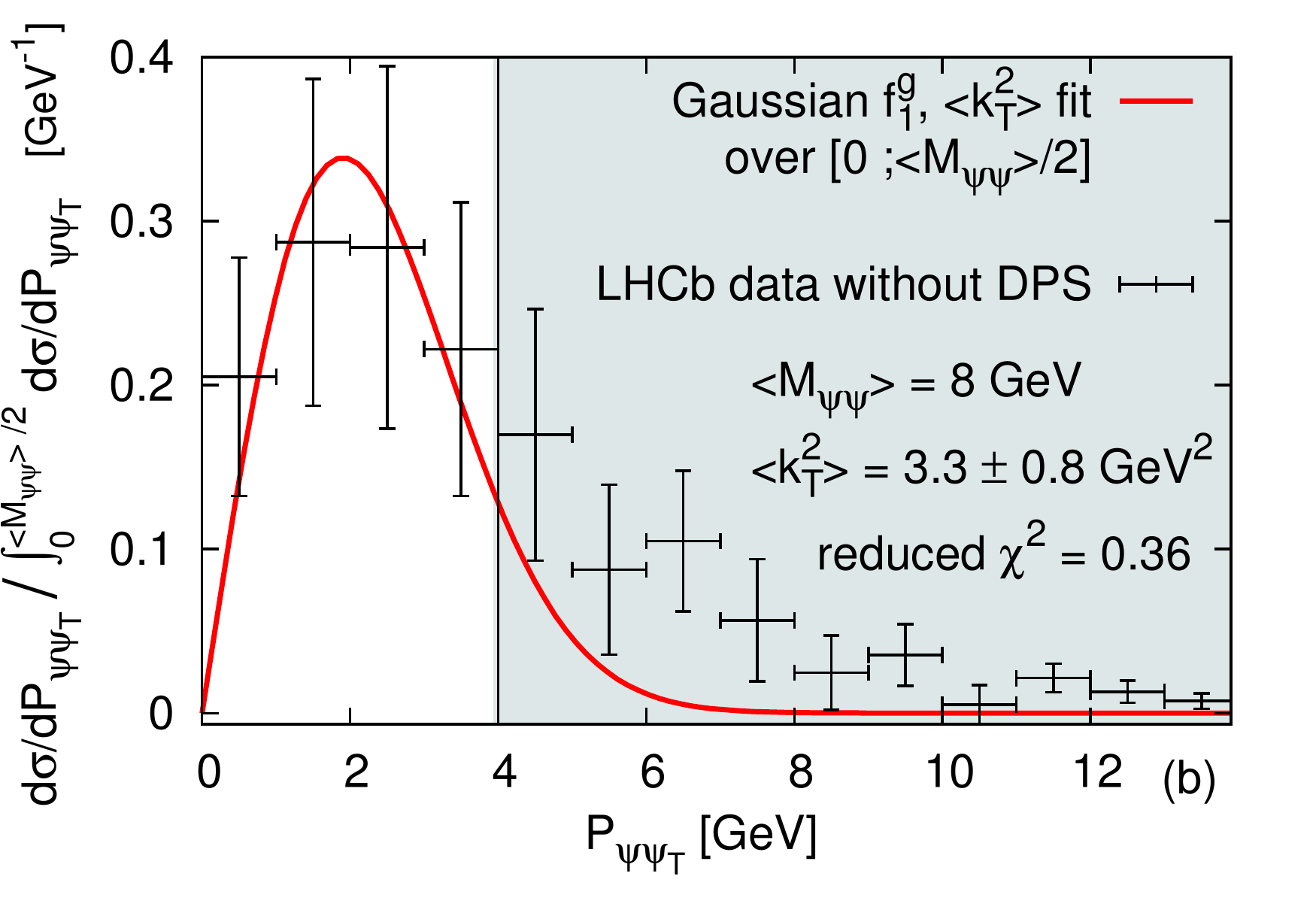}}
\caption{The normalised $\qTnorm$ dependence of the di-$J/\psi$ yield obtained with
a Gaussian $f_1^g$ with $\langle  k_\sT^2 \rangle$ fit to the normalised LHCb data at 13~TeV~\cite{Aaij:2016bqq} [The data in the gray zone were not used for the fit since the TMD framework does not apply there].\label{fig:f1_fit}}
\end{figure}

We further note that, for TMD Ans\"atze with factorised dependences on 
$x$ and $\bm k_\sT^2$, the normalised $\qTnorm$ spectrum depends neither
on $x$  nor on other variables. The data on the $\qTnorm$ spectrum are fitted up to $M_{\Q\Q}/2$, employing a  non-linear least-square minimisation procedure with the LHCb experimental uncertainties used to weight the data. We  obtain $\langle k_\sT^2 \rangle = 3.3 \pm 0.8$~GeV$^2$. The resulting $\chi^2$ is 1.08.

This is the first time that experimental information on gluon TMDs is extracted from a gluon-induced process
with a colourless final state, {for which TMD factorisation should apply.}
The discrepancy between the TMD curve and the data for $\qTnorm \gtrsim M_{\Q\Q}/2$ is expected, 
as it leaves room for hard final-state radiations not accounted for in the TMD approach 
outside of its range of applicability. 

{The data used for our $\langle k_\sT^2 \rangle$ fit correspond to a scale, $\mu$, close to $M_{\Q\Q} \sim 8$~GeV. As such, it should be interpreted as an effective value, including both nonperturbative and perturbative contributions. The latter, through TMD QCD evolution, increases $\langle k_\sT^2 \rangle$ with $\mu$~\cite{Boer:2011kf,Boer:2014tka,Echevarria:2015uaa}. Extracting a genuine nonpertubative $\langle k_\sT^2 \rangle$ [at  $\mu \lesssim 1$~GeV] thus requires to account for TMD evolution along with a fit to data at different scales. Di-$J/\psi$ data from LHCb, CMS and ATLAS should in principle be enough to disentangle these perturbative and nonperturbative evolution effects, yet requiring a careful account for acceptance effects as well as perturbative contributions beyond TMD factorisation; these data are indeed not double differential in $\qTnorm$ and $M_{\psi\psi}$. This is left for a future study.}

{In the above extraction of}  $\langle k_\sT^2 \rangle$, we have neglected the influence of $h_1^{\perp g}$ on
the $\qTnorm$ spectrum. 
The LHCb measurement was made without any transverse-momentum cuts, thus near threshold where $M_{\Q\Q} \sim 2 M_\Q$ and where 
$F_2/F_1$ is close to 0.4~\% (cf. \ce{eq:Fi_threshold}). 
The situation is  analogous to 
$\Q+\gamma$~\cite{Dunnen:2014eta}, $\Q+ \gamma^\star$ or
$\Q+Z$~\cite{Lansberg:2017tlc} with a negligible impact of $h_1^{\perp g}$
on the TM spectra but significantly different from that for  di-photon~\cite{Qiu:2011ai}, {\it single} $\eta_c$~\cite{Boer:2012bt}, di-$\eta_c$\cite{Zhang:2014vmh} and $H^0$+jet~\cite{Boer:2014lka} production. Data nonetheless do not exist yet for any of these channels.
Unfortunately, the CMS di-$\Upsilon$ sample~\cite{Khachatryan:2016ydm} is not large enough (40 events) to
perform a $\langle k_\sT^2 \rangle$ fit at $M_{\Q\Q} \sim 20$~GeV. With 100 fb$^{-1}$ of 13 TeV data, this should be possible.

\section{Azimuthal dependences}\vspace*{-0.2cm}

In the perturbative regime, particularly at large $k_T$, $h_1^{\perp g}$ can be connected~\cite{Boer:2014tka,Echevarria:2015uaa} to $g(x)$ with a $\alpha_s$ pre-factor. In the nonperturbative regime, this connection is lost and we currently do not know whether it is also $\alpha_s$-suppressed. As such, it remains useful to consider} the  model-independent 
positivity bound~\cite{Mulders:2000sh,Cotogno:2017puy}:
\eqs{|h_1^{\perp g}(x,\bm k_\sT^2,\mu)|\le \frac{2M_p^2}{\bm k_\sT^2} f_1^g(x,\bm k_\sT^2,\mu)\label{eq:h1pg-bound}}
holding for any value of  $x$ and $\bm k_\sT^2$.

\begin{figure}[hbt!]
\centering
{\includegraphics[width=\columnwidth]{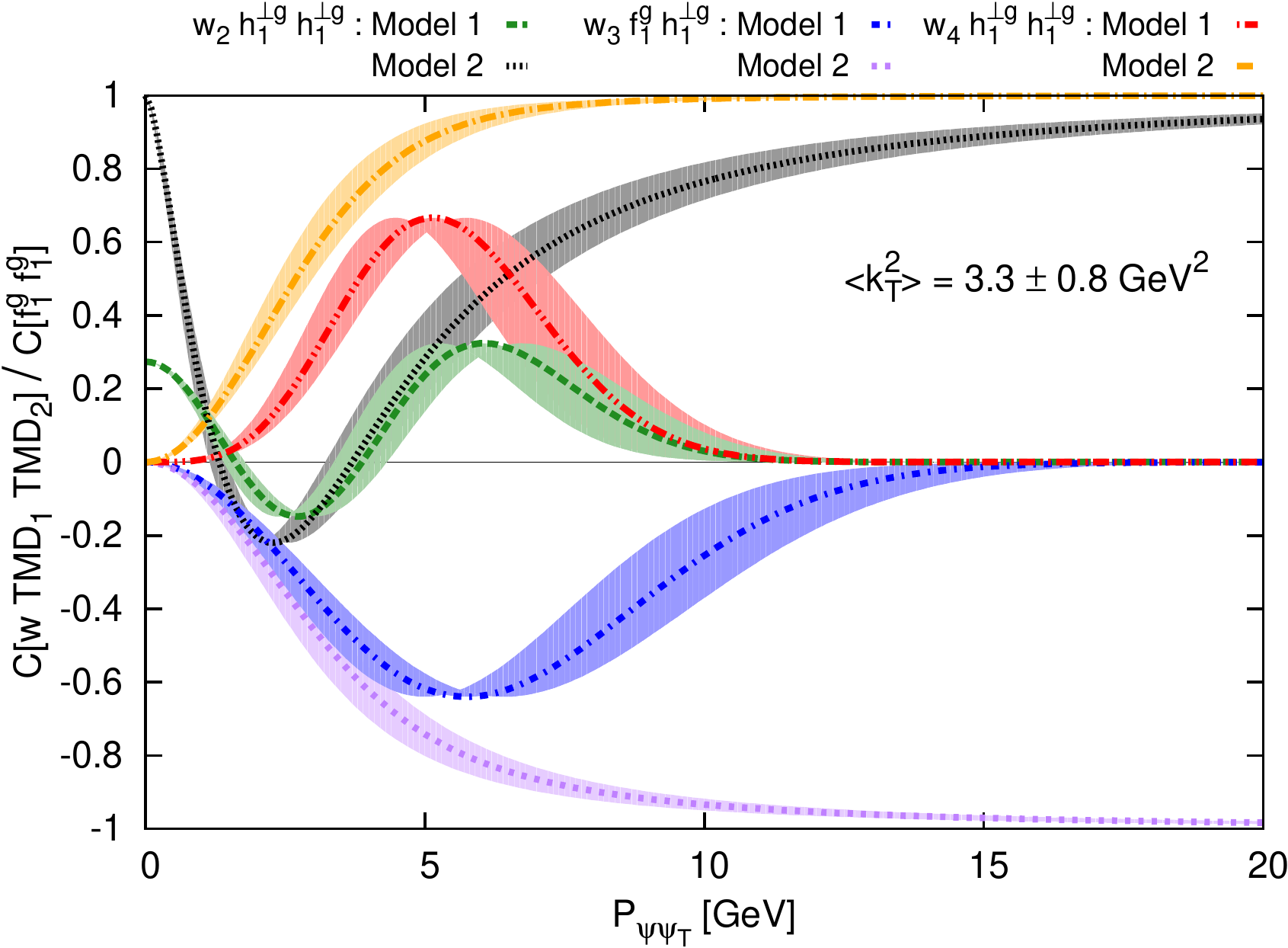}\vspace*{-0.5cm}}
\caption{Various ratios of the TMD convolutions using both our models of $h_1^{\perp g}$
for $\langle  k_\sT^2 \rangle=3.3$~GeV$^2$ (central curves) varied by 0.8~GeV$^2$ (bands).\label{fig:TMD-convolutions}}
\end{figure}

%%%%%%%%%%%%%%%%%%%%%%%%%%%%%%%%%%%%%%%%%%%%%%%%%%%%%%%%%%%%%%%%%%%%%%%%%%%%
\begin{figure*}[hbt!]
\centering
\subfloat{\includegraphics[angle=-90,width=0.48\textwidth]{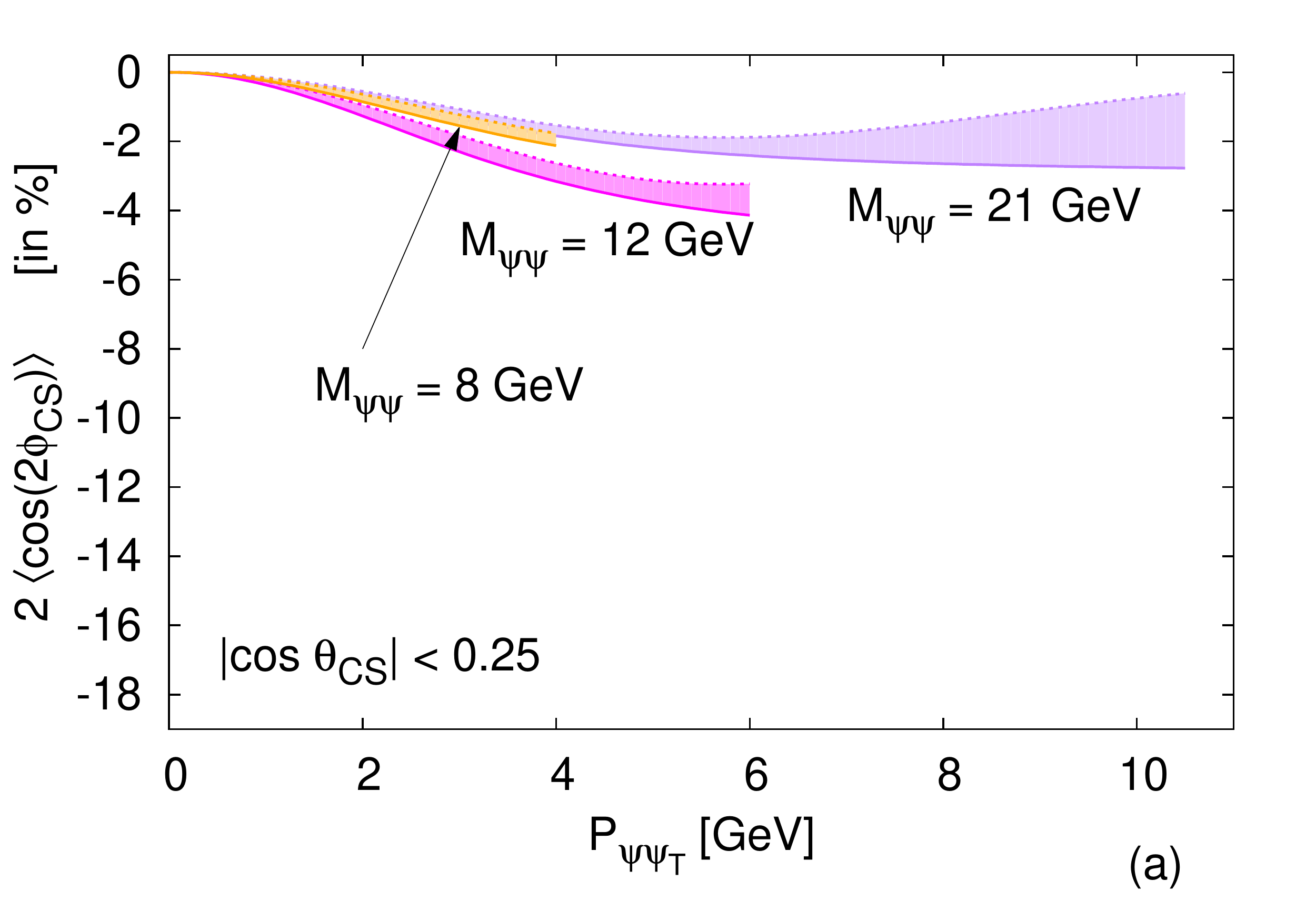}\label{fig:S2_costheta}}\quad
\subfloat{\includegraphics[angle=-90,width=0.48\textwidth]{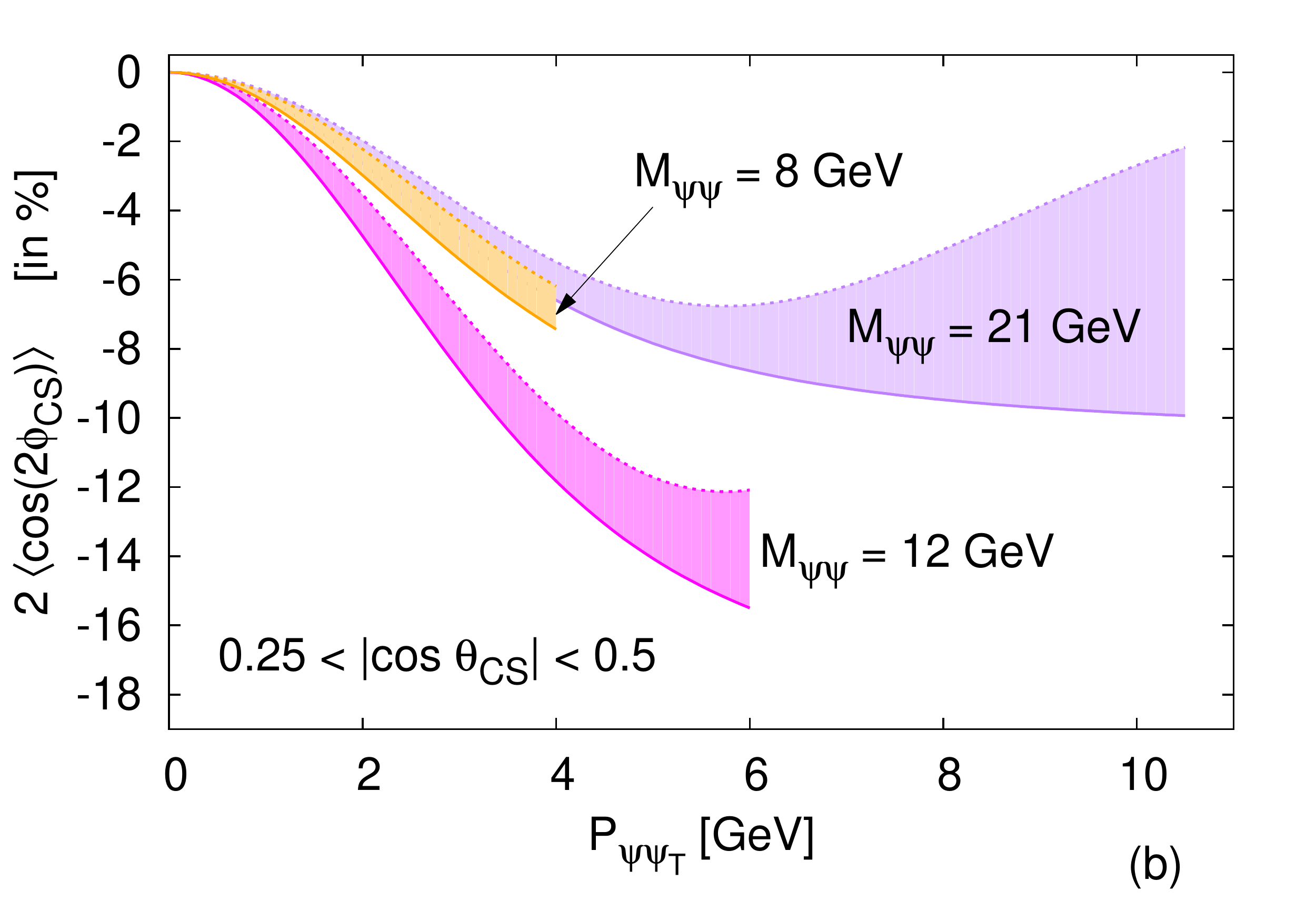}\label{fig:S2_costheta-2}}\vspace*{-0.45cm}\\
\subfloat{\includegraphics[angle=-90,width=0.48\textwidth]{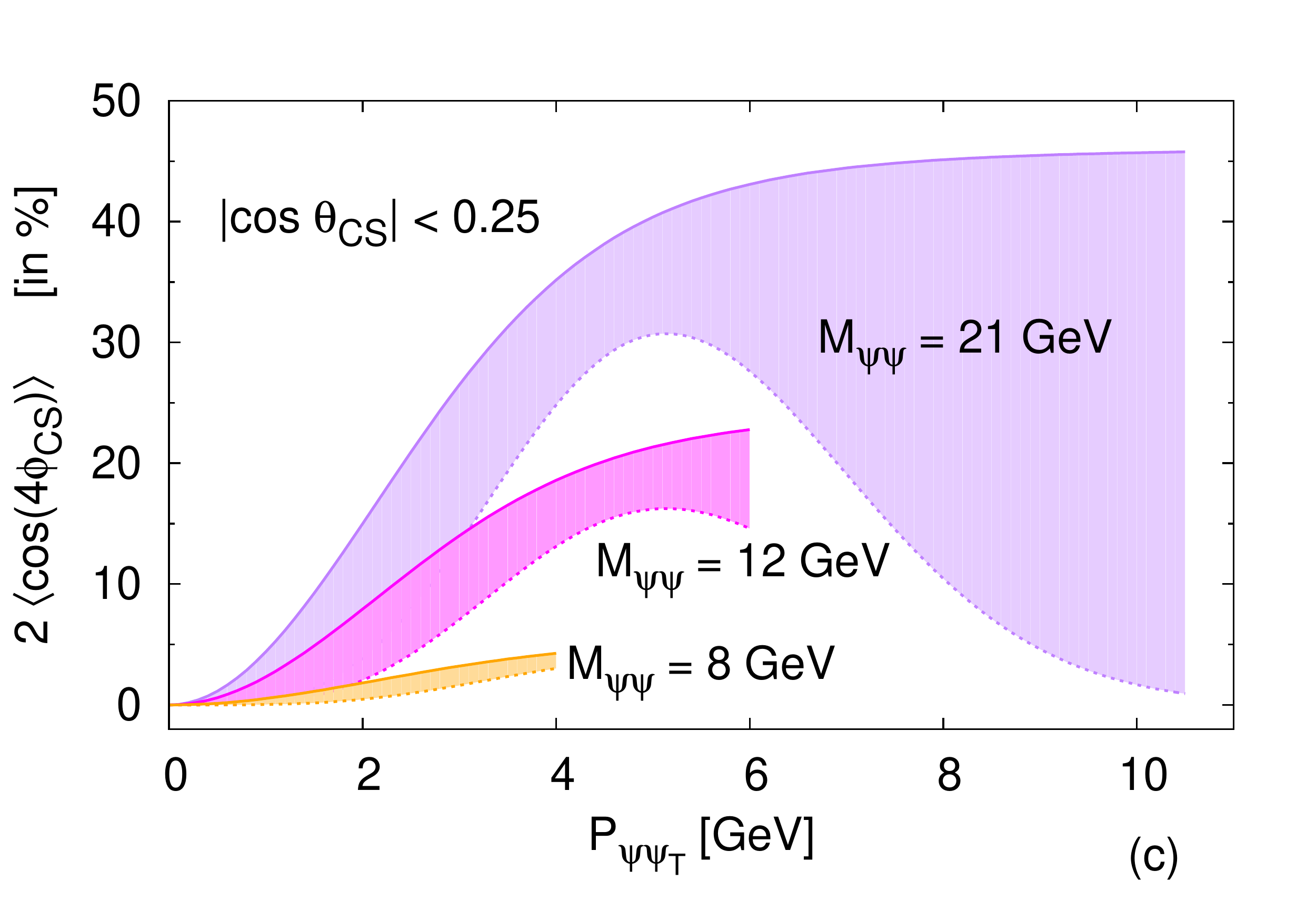}\label{fig:S4_costheta}}\quad
\subfloat{\includegraphics[angle=-90,width=0.48\textwidth]{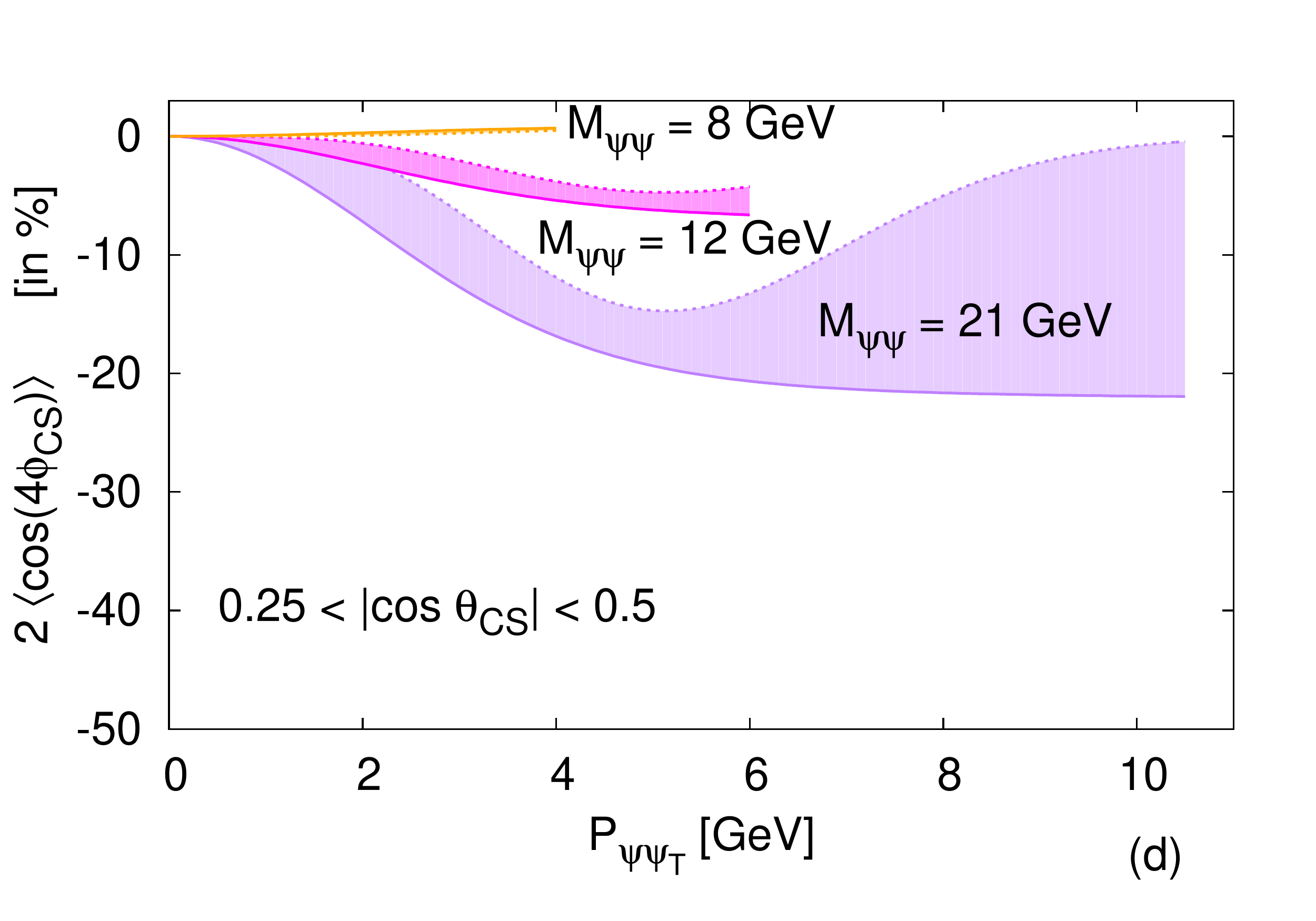}\label{fig:S4_costheta-2}\vspace*{-0.45cm}}
\caption{$2\langle  \cos n\phi_{\CS} \rangle$ for $n=2,4$ computed for $|\cos\theta_{\CS}| < 0.25$ and for $0.25 < \cos \theta_{\rm CS}< 0.5$  for $\langle  k_\sT^2 \rangle=3.3$~GeV$^2$ for 3 values of $M_{\Q\Q}$ (8, 12 and 21 GeV) relevant respectively for the LHCb~\cite{Aaij:2016bqq}, CMS~\cite{Khachatryan:2014iia} and ATLAS~\cite{Aaboud:2016fzt} kinematics. The spectra are plotted up to $M_{\Q\Q}/2$. Our results do not depend on $Y_{\Q\Q}$. The uncertainty bands result from the use of both our models of $h_{1}^{\perp g}$. The solid line, which shows the largest asymmetries corresponds to the Model 2 (saturation of the positivity bound) and the dashed line to Model 1.}
\end{figure*}
%%%%%%%%%%%%%%%%%%%%%%%%%%%%%%%%%%%%%%%%%%%%%%%%%%%%%%%%%%%%%%%%%%%%%%%%%%%

This bound is  satisfied~\cite{Boer:2011kf}  by
\eqs{
h_1^{\perp g}(x,\bm k_\sT^2,\mu)=
 \frac{2 M_p^2}{\langle  k_\sT^2\rangle} \frac{(1-r)}{r}  
\frac{g(x,\mu)}{\pi \langle  k_\sT^2 \rangle} 
\exp\Big(1-\frac{\bm k_\sT^2}{r \langle k_\sT^2 \rangle}\Big)
}
with $r < 1$. We take $r=2/3$ maximising the second
$\kT$ moment   of $h_1^{\perp g}$. We note that such a choice is motivated by previous TMD studies~\cite{Boer:2011kf,Boer:2012bt} where the effects of $h_1^{\perp g}$ were also predicted. In general, values of $r$ smaller than $2/3$ will lead to asymmetries which are narrower in $\qTnorm$, but with a larger maximum. On the other hand, for $r > 2/3$, the asymmetries will be broader and with a smaller peak. With this choice, all 4 TMD convolutions are simple analytical functions whose
$\qTnorm$ dependence is shown on \cf{fig:TMD-convolutions}.
Beside, computations in the high-energy (low-$x$) limit (see \eg\ \cite{Dumitru:2015gaa,Metz:2011wb}) suggest to take
\eqs{
h_1^{\perp g}(x,\bm k_\sT^2,\mu)=\frac{2M_p^2}{\bm k_\sT^2} f_1^g(x,\bm k_\sT^2,\mu).
}

The corresponding convolutions can easily be calculated numerically. Their $\qTnorm$ dependence is  shown 
on \cf{fig:TMD-convolutions} for $\langle  k_\sT^2 \rangle=3.3$~GeV$^2$ (which follows from our fit of $f_1^g$).
As we discuss later, having both these models at hand is very convenient, as it allows us to assess the influence of
the variation of $h_1^{\perp g}$ -- \eg\ due to the scale evolution-- on the observables. "Model 1" will refer to the Gaussian form  with $r=2/3$ and "Model 2" to the form saturating the positivity bound. The  bands in \cf{fig:TMD-convolutions} corresponds to a variation of
$\langle  k_\sT^2 \rangle$ about 3.3 GeV$^2$ by 0.8 GeV$^2$ (which also results from our fit). We note that these bands 
are in general significantly smaller than the difference between the curves for Model 1 and 2.
As such, we will use the results from Model 1 and 2 to derive uncertainty bands which however should remain
indicative since, as stated above, nearly nothing is known about these distributions.

Having fixed the functional form of the TMDs and $\langle k_\sT^2 \rangle$ and having computed
the factors $F_i$, we are now ready to provide predictions
for the azimuthal modulations through $2\langle  \cos n\phi_{\CS} \rangle$
as a function of $\qTnorm$, $\cos\theta_{\CS}$ or $M_{\Q\Q}$. 
\cf{fig:S2_costheta} \& \ref{fig:S4_costheta}  show $2\langle  \cos n\phi_{\CS} \rangle$ 
($n=2,4$) as a function of $\qTnorm$ for both our models of $h_1^{\perp\,g}$ for 3 values of $M_{\Q\Q}$, 8, 12 and 21 GeV for $|\cos\theta_{\CS}| < 0.25$. These values are  relevant respectively 
for the LHCb~\cite{Aaij:2016bqq}, CMS~\cite{Khachatryan:2014iia} and ATLAS~\cite{Aaboud:2016fzt} 
kinematics. Still to keep the TMD description applicable, we have plotted the spectra up to $M_{\Q\Q}/2$ .
Let us also note that with our factorised TMD Ans\"atze, $2\langle  \cos n\phi_{\CS} \rangle$ 
do not depend on $Y_{\Q\Q}$. Indeed, the pair rapidity 
only enters the evaluation of $d\sigma$ via the momentum fractions $x_{1,2}$
 in the TMDs. It thus simplifies in the ratios.

The size of the expected azimuthal asymmetries is
particularly large, \eg\ for $\qTtwo \simeq \langle k_\sT^2 \rangle$. $2\langle  \cos 4\phi_{\CS} \rangle$ even gets close to 50\% in the $\qTnorm$ region probed by CMS and ATLAS for $|\cos\theta_{\CS}| < 0.25$; 
this is probably the highest value ever predicted for a gluon-fusion process which directly
follows from the extremely favourable hard coefficient $F_4$  --as large as $F_1$.
Such values are truly promising to extract the distribution $h_1^{\perp g}$ of linearly-polarised
gluons in the proton which appears quadratically in $2\langle  \cos 4\phi_{\CS} \rangle$. 
In view of these results, it becomes clear that the kinematics of CMS and ATLAS are better suited 
with much larger expected asymmetries than that of LHCb, not far from threshold, unless LHCb
imposes $P_{\psi T}$ cuts.

$2\langle  \cos 2\phi_{\CS} \rangle$ allows one to 
lift the sign degeneracy of $h_1^{\perp g}$ in $2\langle  \cos 4\phi_{\CS} \rangle$ but is below $5\%$ for $|\cos\theta_{\CS}| < 0.25$ (\cf{fig:S2_costheta}). 
This is expected since $F_3$ vanishes for small $\cos\theta_{\CS}$ (\ce{eq:Fi_large_M}). 
It would thus be expedient to extend the range of $|\cos\theta_{\CS}|$
pending the DPS contamination.
Indeed, in view of recent di-$J/\psi$ phenomenological studies~\cite{Lansberg:2014swa,Kom:2011bd,Baranov:2011ch}, one expects the
DPSs to become dominant at large $\Delta y$ while these cannot be treated along the lines of our analysis.
To ensure the SPS dominance, it is thus judicious to avoid 
the region  $\Delta y> 2$, and probably $\Delta y > 1$ to be on the safe side. 
Even though the relation between $\Delta y$ --measured in the
hadronic c.m.s.-- and $\cos\theta_{\CS}$ is in general not trivial, it strongly
simplifies when $P^2_{\Q T} \gg (M^2_\Q,\qTtwo)$, such that $ \cos\theta_{\CS} = 
\tanh{\Delta y/2}$~\footnote{In fact, $\Delta y/2$ then 
coincides with the usual definition of the pseudorapidity of one quarkonium 
since $\Delta y$ is not sensitive to the longitudinal boost between
the $\CS$ frame and the c.m.s.}. Up to  $|\cos\theta_{\CS}| \sim 0.5$, the sample should thus 
remain SPS dominated in particular with the CMS and ATLAS $P_{\Q T}$ cuts.
In fact, in a bin $0.25 <|\cos\theta_{\CS}| < 0.5$, 
$2\langle  \cos 2 \phi_{\CS} \rangle$ nearly reaches 15\% (\cf{fig:S2_costheta-2}). On the contrary, $2\langle  \cos 4\phi_{\CS} \rangle$ exhibits a node 
close to $\cos\theta_{\CS}\sim 0.3$ (\cf{fig:S4_costheta-2}). As such, measuring $2\langle  \cos 4\phi_{\CS} \rangle$ 
for $|\cos\theta_{\CS}| < 0.25$ and $0.25 <|\cos\theta_{\CS}| < 0.5$ would certainly be
instructive. If our models for $h_1^{\perp g}$ are realistic, this is definitely within
the reach of CMS and ATLAS, probably even with data already on tape.

{TMD evolution will affect the size of these asymmetries, although in a hardly quantifiable way. In fact, TMD evolution has never been applied to any $2\to2$ gluon-induced process and is beyond the scope of our analysis. One can however rely on an analogy with a $\eta_b$ production study~\cite{Echevarria:2015uaa} (a $2\to 1$ gluon-induced process
at $\mu \sim 9 $ GeV) where the ratio ${\cal C} [w_2\, h_1^{\perp\,g}\, h_1^{\perp\,g}]/{\cal C}[f_1^g f_1^g] $ was found to range between 0.2 and 0.8. This arises from a subtle interplay between the evolution and the nonperturbative behaviour of $f_1^g$ and $h_1^{\perp\,g}$. We consider that the uncertainty spanned by our Model 1 and 2 gives a fair account of the typical uncertainty of an analysis with TMD evolution, hence the bands in our plots.}

\section{Conclusions\label{sec:conclusion}}\vspace*{-0.2cm}
We have found out that the short-distance coefficients to the azimuthal modulations
of $J/\psi$($\Upsilon$) pair yields equate the azimuthally independent terms, which renders
these processes ideal probes of the linearly-polarised gluon distributions in an unpolarised 
proton, $h_1^{\perp g}$. Experimental data already exist --more will be recorded in the near future-- 
and it only remains to analyse them along the lines discussed above, by evaluating 
the ratios $2\langle  \cos 2\phi_{\CS} \rangle$ and $2\langle  \cos 4\phi_{\CS} \rangle$. In fact, we have already highlighted the relevance of the LHC 
data for di-$J/\psi$ production by constraining, for the first time, the
transverse-momentum dependence of $f_{1}^g$ at a scale close to $2 M_\psi$.

Let us also note that similar measurements can be carried out at fixed-target set-ups where
luminosities are large enough to detect $J/\psi$ pairs. The COMPASS experiment 
with pion beams may also record di-$J/\psi$ events as did NA3 in the 80's~\cite{Badier:1982ae,Badier:1985ri}. Whereas single-$J/\psi$ production may partly be
from quark-antiquark annihilation, di-$J/\psi$ production should mostly be from gluon fusion and thus
analysable along the above discussions. Using the 7 TeV LHC beams~\cite{Lansberg:2015lva} 
in the fixed-target mode with a LHCb-like detector~\cite{Hadjidakis:2018ifr,Massacrier:2015qba,Massacrier:2015nsm,Lansberg:2014myg}, 
one can expect 1000 events per 10 fb$^{-1}$, enough to measure
a possible $x$ dependence of $\kTsqav$ as well as to look for azimuthal asymmetries 
generated by $h_{1}^{\perp g}$. Such analyses could also be complemented with target-spin asymmetry studies~\cite{Kikola:2017hnp,Lansberg:2016gwm,Lansberg:2016urh},
to extract the gluon Sivers function $f_{1T}^{\perp g}$ as well as the gluon transversity distribution $h_{1T}^{g}$ or
the distribution of linearly-polarised gluons in a transversely polarised proton, $h_{1T}^{\perp g}$, paving
the way for an in-depth gluon tomography of the proton.

%%%%%%%%%%%%%%%%%%%%%%%%%%%%%%%%%%%%%%%%%%%%%%%%%%%%%%%%%%%%%%%%%%%%%%%%%%%%

{\bf Acknowledgements.}
We thank A.\ Bacchetta, D.\ Boer, M.\ Echevarria and H.S.\ Shao for useful comments and L.P. Sun for discussions about~\cite{Qiao:2009kg}. 
The work of J.P.L. and F.S. is supported in part by the French IN2P3--CNRS via the LIA FCPPL 
(Quarkonium4AFTER) and the project TMD@NLO. 
The work of C.P. is  supported by the European Research Council (ERC) under the European Union's Horizon 2020 research and innovation program (grant agreement No.\ 647981, 3DSPIN).
The work of M.S. is supported in part by the Bundesministerium f\"ur Bildung
und Forschung (BMBF) grant 05P15VTCA1.

%\end{linenumbers}

\vspace*{-0.05cm}
\appendix

\section{The full expressions of the $f_{i,n}$} \label{sec:appendix}
The factors $f_{i,n}$ are simple polynomials in $\alpha$, \ie\
\begin{align}\label{eq:f1}
&f_{1,0}= 6\alpha^{8}-38\alpha^{6}+83\alpha^{4}+480\alpha^{2}+256, \nn\\
&f_{1,1}=2(1-\alpha^{2})(6\alpha^{8}+159\alpha^{6}-2532\alpha^{4}+884\alpha^{2}+208),\nn \\
&f_{1,2}=2(1-\alpha^{2})^{2}(3\alpha^{8}+19\alpha^{6}+7283\alpha^{4}-8448\alpha^{2}-168),\nn\\
&f_{1,3}=-2(1-\alpha^{2})^{3}(159\alpha^{6}+6944\alpha^{4}-17064\alpha^{2}+3968),\nn\\
&f_{1,4}=(1-\alpha^{2})^{4}(4431\alpha^{4}-27040\alpha^{2}+17824),\nn\\
&f_{1,5}=504(1-\alpha^{2})^{5}(15\alpha^{2}-28), \nn\\
&f_{1,6}=3888(1-\alpha^{2})^{6},\\
%& \nn \\
%\end{align}
%\begin{align}%\label{eq:f2}
&f_{2,0}= \alpha^{4},\nn\\
&f_{2,1}= -2(\alpha^{6}+17\alpha^{4}-126\alpha^{2}+108),\nn\\
&f_{2,2}= (1-\alpha^{2})^{2}(\alpha^{4}+756),\nn\\
&f_{2,3}= -36(1-\alpha^{2})^{3}(\alpha^{2}+24), \nn \\
&f_{2,4}= 324(1-\alpha^{2})^{4}, % \\
%&f_{2,5}= 0,
%f_{2,6}= 0\\
%\end{align}
%\begin{align}%\label{eq:f3}
& \\
&f_{3,0}= \alpha^{2}(16-3\alpha^{2}),\nn\\
&f_{3,1}= 6\alpha^{6}+159\alpha^{4}-1762\alpha^{2}+1584,\nn\\
&f_{3,2}= (1-\alpha^{2})(3\alpha^{6}+19\alpha^{4}+5258\alpha^{2}-6696),\nn\\
&f_{3,3}= -(1-\alpha^{2})^{2}(159\alpha^{4}+5294\alpha^{2}-10584),\nn\\
&f_{3,4}= 18(1-\alpha^{2})^{3}(99\alpha^{2}-412),\nn \\
&f_{3,5}= 1944(1-\alpha^{2})^{4},\\
%&f_{3,6}= 0
%\end{align}
%\begin{align}\label{eq:f4}
%& \nn \\
&f_{4,0}= 3\alpha^{4}-32\alpha^{2}+256, \nn\\
&f_{4,1}= -(6(\alpha^{4}+36\alpha^{2}-756)\alpha^{2}+4768),\nn\\
&f_{4,2}=  3\alpha^{8}+38\alpha^{6}+11994\alpha^{4}-32208\alpha^{2}+20400,\nn\\
&f_{4,3}= -2(1-\alpha^{2})(105\alpha^{6}+5512\alpha^{4}-23120\alpha^{2}+19520),\nn\\
&f_{4,4}= (1-\alpha^{2})^{2}(3459\alpha^{4}-30352\alpha^{2}+38560),\nn\\
&f_{4,5}= 72(1-\alpha^{2})^{3}(105\alpha^{2}-268),\nn \\
&f_{4,6}= 3888(1-\alpha^{2})^{4}.
\end{align}

\bibliographystyle{Science}

\bibliography{TMD-Azim-Psi-Psi-290119}

\clearpage

\end{document}